\documentclass[conference]{IEEEtran}
\IEEEoverridecommandlockouts
\usepackage{cite}
\usepackage{amsmath,amssymb,amsfonts}
\usepackage{algorithmic}
\usepackage{graphicx}
\usepackage{textcomp}
\usepackage{xcolor}
\def\BibTeX{{\rm B\kern-.05em{\sc i\kern-.025em b}\kern-.08em
    T\kern-.1667em\lower.7ex\hbox{E}\kern-.125emX}}
    
\begin{document}

\title{Quantum-Guided Test Case Minimization for LLM-Based Code Generation\\
}

\author{\IEEEauthorblockN{Huixiang Zhang}
\IEEEauthorblockA{\textit{Department of Computer Science} \\
\textit{Lakehead University}\\
Thunder Bay, Canada \\
hzhan102@lakeheadu.ca}
\and
\IEEEauthorblockN{Mahzabeen Emu}
\IEEEauthorblockA{
\textit{Quantum Communications and Computing Research Center} \\
\textit{Department of Electrical and Computer Engineering} \\
\textit{Memorial University of Newfoundland}\\
St. John's, NL, Canada \\
memu@mun.ca}
}
% {\footnotesize \textsuperscript{*}Note: Sub-titles are not captured in Xplore and
% should not be used}
% \thanks{Identify applicable funding agency here. If none, delete this.}
% }

% \author{\IEEEauthorblockN{1\textsuperscript{st} Given Name Surname}
% \IEEEauthorblockA{\textit{dept. name of organization (of Aff.)} \\
% \textit{name of organization (of Aff.)}\\
% City, Country \\
% email address or ORCID}
% \and
% \IEEEauthorblockN{2\textsuperscript{nd} Given Name Surname}
% \IEEEauthorblockA{\textit{dept. name of organization (of Aff.)} \\
% \textit{name of organization (of Aff.)}\\
% City, Country \\
% email address or ORCID}

\maketitle

\begin{abstract}
Precisely controlling Large Language Models (LLMs) to generate efficient and concise code is a central challenge in software engineering. We introduce a framework based on Test-Driven Development (TDD) that transforms code specification into a combinatorial optimization task. The framework first prompts an LLM to generate a test suite, then formulates the Test Case Minimization (TCM) problem as a Quadratic Unconstrained Binary Optimization (QUBO) model. This QUBO paradigm is compatible with both classical solvers and emerging hardware such as quantum annealers. Experimentally, quantum annealing solves the core TCM task 16 times faster than simulated annealing. This performance underpins our end-to-end framework, which reduces total token consumption by 36.5\% and significantly improves code quality. This work demonstrates a powerful synergy between generative AI and combinatorial optimization in software engineering, highlighting the critical importance of precise model formulation.
\end{abstract}


\section{Introduction}
Large Language Models (LLMs) are changing software development. However, their powerful code generation capabilities are often accompanied by a core challenge: a lack of output control. Developers frequently receive code that is functionally correct but unnecessarily redundant and insufficient. The critical task has become how to precisely guide an LLM to produce concise code \cite{schafer2024}.

Test-Driven Development (TDD) offers a paradigm to address this. The test cases serve as a precise and unambiguous guide for the code. A new problem arises when applying TDD to LLMs. Feeding a comprehensive test suite directly into a model incurs prohibitive token and inference costs due to its large size, making it economically and practically infeasible. This creates a central conflict. We need the precision of test cases, but we cannot afford the cost of their completeness \cite{chen2023}.

To resolve this conflict, we position the Test Case Minimization (TCM) problem at the heart of our workflow. We adopted a TCM Quadratic Unconstrained Binary Optimization (QUBO) model for our framework \cite{wang2024}. This provides a powerful mathematical framework and, more importantly, opens the door to acceleration using frontier technologies like quantum computing. In modern software development, where Continuous Integration/Continuous Deployment (CI/CD) is the standard, testing efficiency dictates delivery velocity. A slow test selection process is a major bottleneck. Our work shows that the over 10 times speedup from quantum computing can transform TCM from an infrequent, offline task into a more easy process. This research therefore aims to answer the following central questions:
\begin{itemize}
    \item Can an automated framework combining TDD with combinatorial optimization improve the quality and reduce the cost like token consumption of LLM-generated code?
    \item For the core optimization task within this framework, what is the practical performance advantage of quantum computing over classical algorithms, and what does this advantage mean for software development pipelines?
\end{itemize}

\section{Framework Overview}
Our framework is designed to automatically generate and optimize test suites for code modules. It leverages LLMs for test case generation and employs combinatorial optimization to minimize the size of the test set. The workflow consists of three stages: (1) Comprehensive Test Generation, (2) Test Suite Optimization, and (3) Optional Code Refinement.

The input to this stage is an existing code module that requires testing. Our objective is to generate a comprehensive test suite for this code, where each test has an explicit functional label. We first provide the code under test to an LLM. Using a specifically designed prompt, we instruct the LLM to analyze the code and generate a highly diverse and redundant test suite, denoted as $T_{comprehensive}$. A core requirement of this prompt is that the LLM must assign a clear label to each generated test case, indicating the specific feature that it is designed to validate. This explicit mapping from tests to features is a critical prerequisite for the subsequent optimization stage.

The core task of this stage is to reduce the comprehensive test suite $T_{comprehensive}$ into an efficient, minimal subset $T'$, using the mapping of the test to the characteristic of the previous stage. We formulate this task as a Quadratic Unconstrained Binary Optimization (QUBO) problem. We assign a binary decision variable $t_i$ to each test case in $T_{comprehensive}$, where $t_i=1$ if the test is selected and $0$ otherwise. Our goal is to find an optimal assignment of variables that minimizes a unified objective function.

$$ \min \left( \sum_{i} \text{cost}(t_i) + \lambda \sum_{j} \text{Penalty}_j \right) $$

This objective function intuitively balances two core goals:

\begin{itemize}
    \item \textbf{$\sum_{i} \text{cost}(t_i)$}: This is the cost term. Attempts to minimize the total cost of the selected test cases. The cost can be the number of tests where each $\text{cost}(t_i)$ is 1.

    \item \textbf{$\lambda \cdot \sum_{j} \text{Penalty}_j$}: This is the penalty term. For each functional requirement $j$ that is not covered by any selected test, $\text{Penalty}_j$ incurs a positive penalty value. The coefficient $\lambda$ controls the strength of this penalty.
\end{itemize}

Solving this QUBO model yields a combination of test cases that minimizes the total cost while ensuring that all functional requirements are covered. This optimal combination forms our final minimized test suite, $T'$.

\subsection{Optional Code Refinement}
This optional stage leverages the minimal test suite $T'$ to refine and improve the quality of the original code. We provide the original code and $T'$ to an LLM, instructing it to refactor for efficiency and readability under the strict constraint that all tests must pass.

The minimal test suite $T'$ acts as a precise and unambiguous functional specification in this process. It establishes a clear boundary for the LLM's modifications. This allows the model to safely remove redundant logic or unnecessarily complex implementations from the code without breaking core functionality. The final output of this process is a refined, higher-quality version of the code, term as $C_{\text{final}}$.

\section{Experimental Evaluation}

Our experimental evaluation is designed to answer two core questions: 1) What is the performance of quantum computing versus classical methods on the core task of test suite optimization? 2) What is the impact of our TDD framework on the final output of LLM-based code generation?

\begin{figure}[ht]
    \centering
    \includegraphics[width=0.8\linewidth]{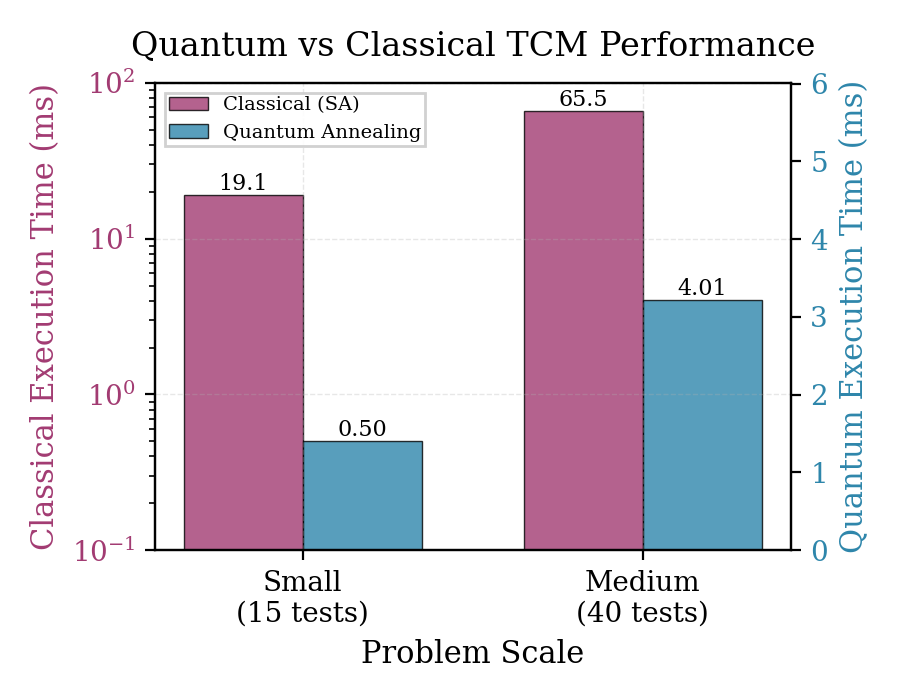}
    \vspace{-4mm}
    \caption{Quantum Annealing are faster then Simulated Annealing for the core TCM optimization task.}
    \label{fig:solver-performance}

\end{figure}

We first benchmarked the core task of our framework's Test Suite Optimization stage. We compared the solving time of a physical quantum annealer against a classical simulated annealing solver on small- and medium-scale TCM problems.

% A deeper analysis revealed that neither the quantum nor the classical solver found a valid solution with 100\% requirement coverage. This was not a failure of the solvers, but a limitation of the QUBO model itself: its static penalty coefficients failed to correctly balance constraints and costs. This finding highlights a crucial insight for practical applications: a high-quality model that guarantees solution validity is the fundamental prerequisite for leveraging any advanced solver, including quantum hardware. Future work will focus on developing dynamic penalty mechanisms.
As shown in Figure 1, the quantum annealer shows a speed advantage. In the medium-scale problem, its resolution time ($4.008$ ms) was more than 16 times faster than that of simulated annealing ($65.5$ ms). This leap in performance is significant. It suggests that the test optimization process can be transformed from a slow, offline batch job into a real-time tool suitable for per-commit execution within a CI/CD pipeline.

% Using the optimized, minimal test suite as a formal specification for LLM code generation delivered significant business value in our end-to-end evaluation. As shown in Table, our TDD-guided framework reduced total token consumption by \textbf{36.5\%} (lowering API costs), decreased code complexity by \textbf{26.1\%} (improving maintainability), and cut the overall generation time by \textbf{54.3\%} (accelerating the development cycle). This proves that providing precise, optimized inputs is key to achieving high-quality, cost-effective AIGC implementation.

Next, we evaluated the impact of our framework on the LLM code generation task. We defined two modes:
\begin{itemize}
    \item \textbf{Baseline:} The original full test suite generated by the LLM is used directly as the specification to guide the generation of the code.
    \item \textbf{TDD-Guided:} Our framework is used to first find a minimal test suite via QUBO optimization, which then guides the code generation.
\end{itemize}

We assessed the final generated code in three dimensions: total token consumption or cost, code quality, and generation time. For code quality, we used Cyclomatic Complexity, an industry-standard metric that evaluates code complexity by counting the number of linearly independent paths. A lower score indicates simpler, more maintainable code.

\begin{table}[h]
\centering
\caption{Performance Comparison: Baseline vs TDD-Guided}
\begin{tabular}{lcccc}
\hline
\textbf{Metric} & \textbf{Baseline} & \textbf{TDD} & \textbf{Improvement} \\
\hline
Total Tokens & 897.75 & 570.25 & \textbf{36.5\%} \\
Complexity Score & 27.75 & 20.50 & \textbf{26.1\%} \\
Generation Time (s) & 40.23 & 18.38 & \textbf{54.3\%} \\
\hline
\end{tabular}
\end{table}

The results, presented in Table 1, show that our TDD-Guided approach has improvements across all metrics:
\begin{itemize}
    \item \textbf{Cost Reduction:} Total tokens were reduced by \textbf{36.5\%}, which translates directly to lower API costs and reduced computational requirements.
    \item \textbf{Quality Improvement:} The code's cyclomatic complexity was lowered by \textbf{26.1\%}, providing quantitative evidence that the code generated from a minimal test set is structurally simpler and of higher quality.
    \item \textbf{Efficiency Gain:} The generation time was shortened by \textbf{54.3\%}. This demonstrates that despite the added optimization step, providing the LLM with a highly focused and concise input dramatically reduces its inference burden, leading to a faster overall process.
\end{itemize}

\bibliographystyle{IEEEtran}
\bibliography{tcm}

\end{document}